\documentclass{article}
\usepackage{hyperref}
\usepackage{amscd,amsmath,amsthm,amssymb}
\usepackage{enumerate,varioref}
\usepackage{epsfig}
\usepackage{graphicx}
\usepackage{subfig}
\usepackage{mathtools}
\usepackage{tikz}
\usepackage{algorithmicx}
\usepackage{algorithm}
\usepackage{algpseudocode}
\usepackage{xcolor}
\newtheorem{thm}{Theorem}

\theoremstyle{definition}
\newtheorem{defn}[thm]{Definition}
\theoremstyle{remark}

\newcommand{\R}{\mathbb{R}}

\title{Portfolio Optimization of 40 Stocks Using DWave’s Quantum Annealer}
\author{Chicago Quantum\footnote{Jeffrey Cohen, Alex Khan, Clark Alexander} \\ email \href{mailto:research@quantum-usaci.com}{the authors}}

\begin{document}

\maketitle

\tableofcontents

\begin{abstract}
	We investigate the use of quantum computers for building a portfolio out of a universe of U.S. listed, liquid equities that contains an optimal set of stocks. Starting from historical market data, we look at various problem formulations on the D-Wave Systems Inc. D-Wave 2000Q$\textsuperscript{TM}$ System (hereafter called DWave)  to find the optimal risk vs return portfolio; an optimized portfolio based on the Markowitz formulation and the Sharpe ratio, a simplified Chicago quantum ratio (CQR), then a new Chicago quantum net score (CQNS). We approach this first classically, then by our new method on DWave. Our results show that practitioners can use a DWave to select attractive portfolios out of 40 U.S. liquid equities.  
\end{abstract}

\section{Introduction}

The challenge we approach in financial portfolio optimization is to maximize expected returns while minimizing variability of expected returns, or risk.  This is a buy and hold strategy and not a mid or high frequency trading strategy.  It relies on previous period risk, in our case one year of daily adjusted close data, and the underlying variability and relationships of equities.  We believe investors can improve their chances by selecting the right combination of stocks.

Among the major challenges in financial portfolio optimization is ``how does an investor balance long term investments between expected return and volatility?"  In this work we tackle this question from a variety of methods.This problem is particularly well suited for an annealing solution, either classical simulated thermal annealing, or quantum annealing since we wish to consider $N$ equities in which each equity may be included in a portfolio or not.  This yields exactly $2^N$ possibilities.  For a potential list of equities as small as 40, this becomes nearly infeasible on a workstation.  When we approach the entirety of the S\&P 500, we very quickly run into a solution space which is computationally infinite.  That is, we do not have enough memory in the observable universe to run through a brute force solution.  

This work is structured as follows: In $\S$2 we begin our exploration with the Sharpe ratio
\begin{equation}
S_a(w) = \frac{w\beta\mathbb{E}[R_a- R_b] + R_b}{\sigma_a} \label{Sharpe}
\end{equation}
Where $\beta$ is the ratio of Covariance of a portfolio with the market over the variance of the entire market \cite{BETA}, $R_a$ is the return of the collection of assets, $R_b$ is the risk free return, and $\sigma_a$ is the standard deviation of the collection of assets, and $w$ is a vector of weights for assets in our portfolio.

We can also see the Sharpe ratio in matrix form as 

\begin{equation}
S_a(w) = \frac{w\beta\mathbb{E}[R_a-R_b] + R_b}{\left[w^t \text{Cov}_{ij} w \right]^{1/2}} \label{sharpematrix}
\end{equation}

From here we develop the Chicago Quantum Ratio (CQR)
\begin{equation}
\text{CQR}_a(w) = \frac{w\cdot \text{Cov}_{im}}{\sigma_a} 
\end{equation}
where $\text{Cov}_{im}$ is the covariance of the $i^{th}$ asset against the entire market.  This is a slight improvement over the Sharpe ratio in terms of computation as we need not consider nominal assets.  Risk free investments have a near zero covariance with the entire market. 

We can also reformulate CQR in matrix form as
\begin{equation}
\text{CQR}_a(w) = \frac{w\cdot \text{Cov}_{im}}{\left[w^t \text{Cov}_{ij} w \right]^{1/2}} \label{CQRMatrix}
\end{equation}

We explore these formulations by a variety of classical methods which one will find in $\S$3. Both formulations are ratios and thus neither is properly suitable for a quantum annealing solution, as DWave requires a linear quadratic form.  We attempt to rectify this by exploring

\begin{equation}
\ln(S_a) = \ln(\mathbb{E}[R_a - R_b]) - \ln(\sigma_a)
\end{equation}

This, however causes a different set of mathematical problems in formulating a consistent quadratic form.  Finally we settle on the Chicago Quantum Net Score (CQNS) which is given by
\begin{equation}
\text{CQNS}(w;\alpha) = Var(R_w) - \mathbb{E}[R_w]^{2+\alpha}
\end{equation}
Where $R_w$ is a weighted portfolio and $\alpha \in\R$ In most experiments we choose an equal weighting i.e. $w_i = 1/n$ where $n$ is the number of assets included, and we choose $\alpha$ near 1.  These are not requirements, but they do make the computations on DWave slightly easier. There is a wide open question as to finding optimal weighting and optimal $\alpha$.

We explain how to formulate a quadratic form for use on DWave in $\S$4.

Finally in $\S$5 and 6 we give our results visually and mathematically, and discuss our future work.

\section{Validity of the Formulation}

\begin{figure}[!ht]
	\centering
	\includegraphics[width=0.9\textwidth]{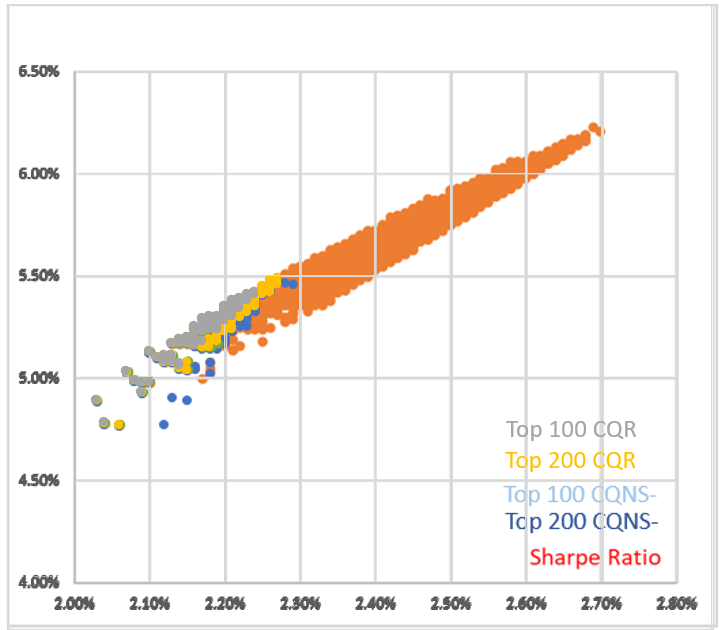}
	\caption{Comparison of CQR and CQNS scores against the Sharpe Ratio}
	\label{fig: Efficient Frontier}
\end{figure}

In its current capacity DWave solves problems which are formulated in terms of an Ising model.  Thus our practical challenge is to provide for Dwave an acceptable model on which it may begin its computations. Consider the following image \ref{fig: Efficient Frontier}

Our formulation has a propensity toward conservative side in investment terms, however, is also demonstrates that the present formulations are near the efficient frontier of investment portfolios.  Thus from an empirical perspective this formulation passes muster.  We develop the method in more detail in $\S$4.

\section{Classical Methods}

In this section we wish to give various formulations which we run on digital computers.  These methods are meant as benchmarking measures so that we may check whether the computations from the annealing quantum computer are legitimate or if the annealing computer lands on local minima which are not particularly deep.

\subsection{Brute Force}
For a smaller asset universe we are able to simply loop through all binary solutions.  If given enough time one can brute force roughly 40 assets.  This however is best approached by an in-place in-time algorithm.  If one attempts to build all $2^N$ portfolios first and simply loop through a list, one will run out of memory.  An in-place algorithm eliminates the memory excess.  An in-time algorithm allows us to write out solutions in case of interruption.  With brute force methods we can explicitly know the minimum possible energy level and thus verify whether our formulation for an annealing quantum computer is valid.

\subsection{Genetic Algorithm}

Our genetic algorithm solution gets to a local minimum deeper than our Monte Carlo method with 950M samples, and does so very quickly.  Our difficulty is in tuning the parameters for number of evolutionary steps, probability of elitism, and size of initial population.  Even with essentially random guesses at these parameters, our genetic algorithm reaches a low enough energy level so that we can determine whether the quantum annealing solutions are legitimate.  

\subsection{Random Sampling}

As mentioned earlier $\S$3.1 a 40 asset portfolio is slightly more than a workstation can handle without an in-place algorithm.  Thus we randomly sample as much as we can.  We are able to sample roughly $2^{29}$ portfolios of a potential $2^{40}$  This means most of our effort is spent around portfolios of size $40 \pm \sqrt{40}$.  Percentage wise this doesn't cover much of the entire spectrum, but we approach $0.4\%$ of the mid sized portfolio.  Specifically by Stirlong's approximation we know.
\[
P(|X|=20) = \frac{1}{2^{40}}\binom{40}{20} \approx \frac{1}{\sqrt{20\pi}}\approx 0.12
\]
On the other hand $2^{29}/2^{40} \approx 0.0005$.  So we get about an 8 fold lift around the middle portfolios.

\subsection{Heuristic Approach}

After running a number of the previous classical methodologies we notice that certain stocks appear most often within the best performing portfolios.  We name these stocks ``All stars."  Similarly we notice several stocks which appear most often in the worst performing portfolios.  We name these ``Dog stars."  The heuristic approach is to attempt building portfolios of mostly All stars with the addition of a few extra items.  This works well as a seeding algorithm for other probabilistic methods and will inform our approach when we attempt to solve the portfolio optimization problem on quantum circuit and trapped ion models.

\subsection{Simulated Annealer as a Monte Carlo}
The original test of our problem comes in the form of a simulated (thermal) annealing solution.  Using statistics of random matrices we are able to tune the parameters of our simulated annealing solution to deliver very deep local minima.  Additionally, this style solution only covers minimizing risk in a portfolio. Based on the cooling rate of the thermal annealing and the number of attempts per solution we use simple statistics from sampling theory to provide a measure of goodness.

\section{Using an Annealing Quantum Computer}

\subsection{The Optimal Portfolio}
The optimal portfolio in our case is one which maximizes the Sharpe ratio.  However, as presented the Sharpe ratio of a portfolio is not computable as a QUBO.  The main thrust of this research is, in fact, how to formulate a QUBO which, when presented to DWave produces similar results to the classical Sharpe ratio.  Consider the following, The Sharpe ratio is defined above in \ref{Sharpe} 
\[
S_a  = \frac{\beta\mathbb{E}[R_a- R_b] + R_b}{\sigma_a}
\]
The numerator can be expressed as a simple dot product, i.e. \[\sum \mu_i w_i\] where $\mu_i$ and $w_i$ are the expected return and the relative weight of the $i^{th}$ asset, respectively.  The denominator can be expressed as the square root of a quadratic form, i.e.
\[
\left[ \frac{1}{|\mathcal{U}|^2}\sum v_i q_i  + 2 \sum_{i<j} cov_{ij}q_i q_j \right]^{1/2}
\]

Where $q_i \in \{0,1\}$ is a binary classifying whether the $i^{th}$ asset is in the portfolio or not, $v_i$ is the variance, and $cov_{ij}$ is the covariance term between asset $i$ and $j$.  One will immediately recognize this as $\sigma_a$ as in the initial formula. One will also recognize that the Sharpe ratio is not a proper quadratic form, and thus not suitable for DWave in its current iteration.  We find that the Chicago Quantum Net Score (CQNS) solves this problem and can be presented as a quadratic form.

\subsection{Developing the QUBO to Number of Assets in a Portfolio}
Consider a universe $\mathcal{U}$ of $N$ assets. When dealing with a single asset portfolio, we only consider the linear terms in a QUBO.  In particular when we have a lower triangular matrix or a zero diagonal matrix, products of the form 
\[
e_i^tQ e_i = 0 
\]

Thus we pick off only the linear terms. In this case, we concisely model the inverse Sharpe Ratio on qubits and use a penalty on the couplers.  DWave finds one-asset portfolios with the highest ratios.

Moving to two or more assets we have substantially more work to do. Looking at a single asset, there are no covariance terms to deal with, and we can embed the inverse Sharpe ratio directly onto the qubits.  We create a unique QUBO for each size portfolio evaluated $\{2,\dots,N\}$ by applying the weights directly to the matrix, so $q_i$ and $q_j$ can remain binary. We divide the linear terms by $N$ ($\sum w_i = 1$) and apply the linear affine transformation.
We divide the variance terms (the diagonal entries) by $N^2 * (N-1)$ to avoid duplication, and divide the covariance terms (off diagonal entries) by $N^2$ to avoid duplication.  We then apply the quadratic affine transformation.
Then we assemble the matrix and reverse the sign on the linear terms.
Finally, we apply a scale factor $(-1,1)$ to the QUBO and write it into our $N\times N\times N$ matrix for processing by DWave

\subsection{Embedding, Scaling, and Hardware Considerations}

We embed the CQNS on DWave by writing the expected returns onto the linear terms, both variance (diagonal) and covariance (off diagonal) onto the quadratic terms. From here the DWave inspector shows how the system encodes and embeds assets onto physical qubits.  An attempt at changing the formulation by manually embedding terms to respect a reordering of assets  does not yield substantial improvements in performance thus we use DWave’s automated embedding functions. 

As we increase our asset size, we see that for a fully connected QUBO DWave requires multiple qubits in chains to leverage the available connections in other groups of the chimera structure.  Increasing our portfolio size above 40 would result in increasing qubit counts utilized to support multiple chains, and the potential for increased chain breaks. We see consistent results with 40 assets when we tune the ``Chain Strength” and scale factors. 

We attempt to reduce our resource cost by removing links between assets that are thought to be insignificant due to low correlation values in our calculations. However, we create inaccurate results in this formulation and from this point we shall avoid this method of reducing our resource cost. 

We adjust DWave’s chain strength parameter to see adequate results with low volume of chain breaks. DWave defaults to a chain strength of 1, and we found we could reduce chain breaks by setting chain strength to values as high as 15.  However, a chain strength of 1 provides more valid answers to our particular QUBO.

We control the qubit value scaling to avoid unequal scaling of linear and quadratic terms. Dwave's native scaling, if left untouched would reduce the accuracy of our scores.  We scale the values we send to DWave within the QUBO.  The process of scaling naturally moves us toward scaling our values by a hyperbolic tangent.  In particular, we find the original scale of our values has some very small values owing to covariance matrices having a zero eigenvalue.  The hyperbolic tangent scales our values to the range $(-1,1)$ we cutoff at $\pm0.99$.  The original scaling produced results which are difficult to read and slightly inconsistent. The newer scaling gives more reliable and consistent results.

\subsection{Affine Transformations of the QUBO}

When exploring portfolios of different sizes, we present a different matrix to Dwave for each desired size of portfolio. We add a penalty for exploring portfolios of different sizes, while maintaining accurate values for the desired portfolio size. The intuition for this follows closely from converting a QUBO into an Ising model.  In order to convert a QUBO into an Ising model we consider the transformation on the binary vector $x$:
\[
z = 2x - 1
\]

This transforms $x^t Qx$ into $z^t J z + c\cdot z + k$ where $c$ is a vector of matching length and $k$ is a constant which we can remove from consideration.  Since we're only looking for the \emph{location} of the lowest energy level in $z$ coordinates, we convert back to $x$ and find the actual lowest energy.  Thus our intution leads us to consider affine transformations in $x$:
\begin{eqnarray}
z & = & ax + b\\
& \implies & x^t Q x \nonumber \\
& = & z^t J z + c\cdot z + k
\end{eqnarray}

where $J = Q/a^2$, $c = -2J\cdot b$ and again $k$ is a constant about which we care none.  Our goal at this point is to find a \emph{shift} which we can apply to the quadratic form.  This corresponds to a translation and is thus closely related to the term $b$ above.  In our formulation we do not use \emph{balance} although our mathematics takes it into consideration.  In this case balance will correspond to a boost in our coordinate system.  

In order to make things easier from the point of view of computation we do not give explicit formulations of shift and balance in terms of $a$ and $b$ above, but rather explain what we actually compute.
\begin{defn}
	Given a universe $\mathcal{U}$ of assets from which to choose, we define the shift factor $s_n$ for exactly $n > 1$ assets from $\mathcal{U}$ as
	\begin{equation}
	s_n = \frac{-2gnm}{|\mathcal{U}|}
	\end{equation}
	where $g$ is the best score derived from a classical simulation in this case a genetic algorithm, $1.5 < m < 20$ is a multiplier which we derive empirically. 
\end{defn}

Our multiplier $m$ is generally around 5. Intuitively, errors can be multiplicative and so we consider values close the geometric average of 1.5 and 20 i.e. $\sqrt{30} \approx 5.4$


As mentioned we skip the algebraic coordinate transformation and simply add our shift factor to both linear and quadratic terms, but we do so as follows:

\begin{itemize}
	\item[1.] to linear terms add $s_n/n$ that is $-gm/|\mathcal{U}|$
	\item[2.] to quadratic terms add $2s_n/(n-1)$
\end{itemize}

\subsection{Visualization}

Visualization of the energy landscape is critical in learning how DWave finds a solution.  It also aids our understanding of how matrix transformations adjust the landscape to improve the probability of sampling ``correct" asset values.  For example, we can place a penalty on smaller portfolios by adding a shift term while subtracting a shift term places a penalty on larger portfolios. Consider the following image:

\begin{figure}[h!]
	\includegraphics[width=0.9\textwidth]{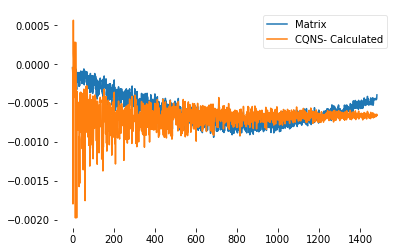}
	\caption{CQNS scores raw vs computed with a shifted matrix}
	\label{fig: Shifting Matrix}
\end{figure}

The values in both of these graphs are computed energy values and the x-axis is the set of assets sorted by number of assets in a portfolio.  We see an unevenly shifted matrix where we have chosen exactly half the possible assets.  We achieve a nice ``U"-shaped curve where the minimal values are clustered around $|\mathcal{U}|/2$ assets.  Furthermore, around $|\mathcal{U}|/2$ our shifted matrix gives us lower values than the raw CQNS. We repeat this for each number of assets $n>1$, while holding the raw CQNS scores for the desired portfolio size constant.  We `tilt’ the curve toward or away from smaller portfolios, with the opposite impact on larger ones, to ensure DWave finds the desired portfolio sizes in that QUBO.

\section{Results}

Our experimental workflow is as follows:
\begin{enumerate}
	\item Download 1-year of daily market data for a specific set of N assets and indices. 
	\begin{itemize}
		\item[$\cdot$] Current as of that moment
		\item[$\cdot$] Hold that data for all experiments
	\end{itemize}
	\item Calculate covariance of each asset with the market, and $\beta$,\cite{BETA} based on log returns
	\item Calculate covariance terms between assets
	\item Calculate underlying and summary values, including Sharpe Ratio and Chicago Quantum Net Score, for an all asset portfolio (i.e. hold all 40 assets for an equal investment amount)
	\item Derive a QUBO for each portfolio size (2 to $|\mathcal{U}|$).
	\begin{itemize}
		\item[$\cdot$] Visualize minimum CQNS values on multiple QUBO matrices.
		\item[$\cdot$] Shift each QUBO to increase likelihood of choosing a portfolio with fixed number of assets.
	\end{itemize}
	\item Run a classical probabilistic algorithm, in our case a genetic algorithm, to see one ``best" portfolio and its values.
	\item Execute DWave using appropriate range of portfolio sizes.
	\item Use the Dwave results the seed the genetic algorithm
	\item Compare values to classical methods.    
\end{enumerate}

The following figures \ref{fig: dwave1}\ref{fig:CQR/CQNS} give some idea of how well the Quantum computer performs using the CQNS against the Sharpe ratio.  We see that in this sample, DWave approaches the efficient frontier in a few cases (highest return for that level of risk).  Most points achieve relative parity with random sampling results, and in some cases DWave suggests lower performing portfolios. These results vary by sample, sample size, and market conditions.  One will also notice that toward the lower left of the efficient frontier, there is a higher density of solutions which shows us that the CQNS formulation lands on the efficient frontier, but is somewhat more conservative.  In reality, the CQNS tends to favor portfolios with lower risk.

\begin{figure}[!ht]
	\begin{center}
		\includegraphics[width=0.75\textwidth]{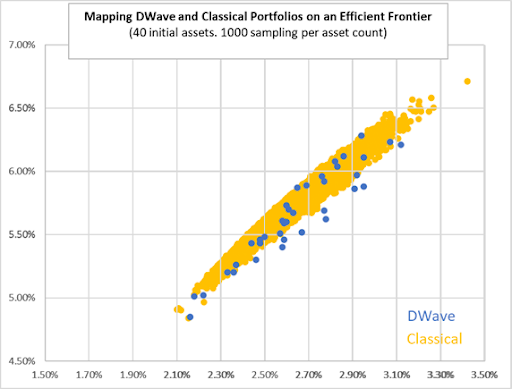}
		\caption{DWave solutions; expected return vs standard deviation.  We also plot Expected return vs Market Momentum, which is a covariance with the market without adjusting for nominal returns as in the Sharpe Ratio.  
		}
		\label{fig: dwave1}
	\end{center}
\end{figure}

\begin{figure}[!ht]
	\centering
	\subfloat[CQR vs DWave]{{\includegraphics[width=5.5cm]{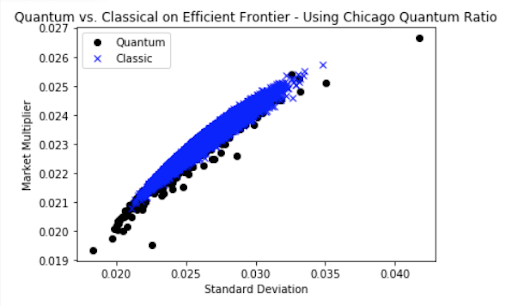} }}%
	\qquad
	\subfloat[CQNS vs DWave]{{\includegraphics[width=5.5cm]{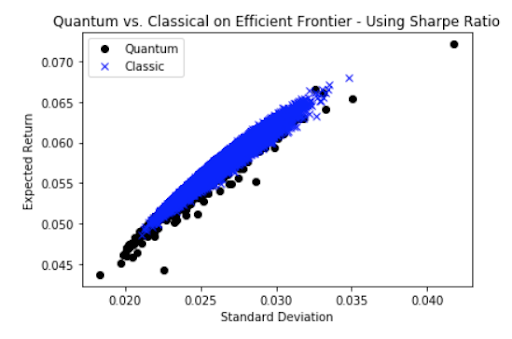} }}%
	\caption{Some plots of Classical vs Quantum annealer computations}%
	\label{fig:CQR/CQNS}%
\end{figure}

We further give results of the CQNS by method.  One will notice that DWave performs well, in fact obtaining better results than Monte Carlo methods, but underperforming the genetic algorithms.  Using DWave as a seed guarantees the genetic algorithm will perform better, as we disallow anything ``worse" than the seed to propagate through generations of solutions.  Interestingly, the two genetic algorithms give the same answers.  We were lucky, however to be able to run the genetic algorithms through many generations.  We expect that if our universe were to have $\sim 1024$ assets the genetic algorithm with DWave seeds would be the top performer, with the simple genetic algorithm performing close to DWave.  

\begin{figure}[!ht]
	\centering
	\subfloat[CQNS by Method]{{\includegraphics[width=5.5cm]{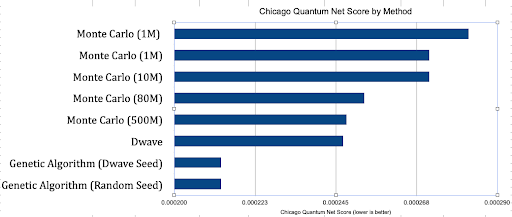} }}%
	\qquad
	\subfloat[Completion Time by Method]{{\includegraphics[width=5.5cm]{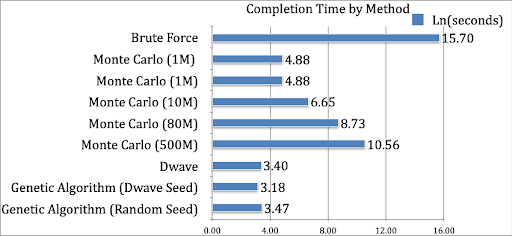} }}%
	\caption{Some plots of Classical vs Quantum annealer computations}%
	\label{fig:Methods}%
\end{figure}

We see that DWave outperforms classical random sampling on average, for all portfolios $n \in \{3-25,27,28,33\}$, which is where most of our portfolios were run.  This shows that DWave is not picking randomly or average solutions, but good ones. The under performance for larger portfolios gives us food for thought in future experiments.

\begin{figure}[!ht]
	\includegraphics[width=0.9\textwidth]{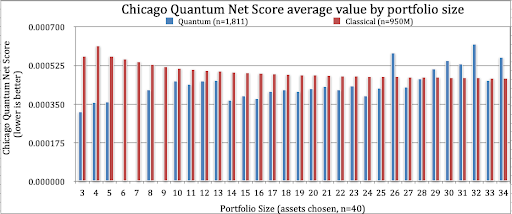}
	\caption{CQNS by asset size}
	\label{fig: n assets}
\end{figure}

\section{Discussion and Conclusion}

\paragraph{Positive Semi-Definite Considerations} 
Practitioners of numpy will know well that numpy is prone to rounding errors. In particular we find that numpy computes covariance matrices with slightly negative eigenvalues $\sim -1e-8$.  While this is not a particularly large negative value, it does open the possibility of a ``minimum risk portfolio" by having a negative overall variance, then the computed Sharpe ratio will be several orders of magnitude too large.  In order to mitigate this we first test our covariance with a Cholesky decomposition.
Second we can simply modify our matrix by computing the standard eigendecomposition and setting all eigenvalues below some absolute threshold at exactly zero.

\paragraph{Considerations of Weighting Assets}
In order to maintain a weighting of assets which sums to 1, we restrict from negative values.  In principle one can short assets, but designing a quadratic form to account for this is a separate problem.  In particular if we short one asset then the sum of positive investments will be greater than one and incurs questions about over and under leveraging. This is out of the scope of our present research. The second problem is that we will not have a QUBO as we must consider values $\{-1,0,1\}$.  This requires a tertiary optimizer, not a binary one.  For this research we have chosen an even positive weighting of assets to reduce computation space.  When we have $|\mathcal{U}|$ assets, our search space with even weightings is $2^{|\mathcal{U}|}$.  Were we to allow a continuous weighting, we would have to approach this with a different optimization scheme.  If we were to allow assets to have a discrete weighting, e.g. 0 to 1 by 0.01, our search space becomes $101^{|\mathcal{U}|}$ which is approximately 7 order of magnitude larger.  This brings our ability to search for an optimal portfolio down to only 32.  We demonstrated that a 32 asset portfolio can be optimized reasonably well with classical methods.  Specifically, we have enough solution space to line up eigenvectors much more easily.  

\paragraph{Selecting our Economic Values}

We use three market equity indices to derive our one-year market returns (Wilshire 5000, S\&P 500, Russell 2000), along with the average return over the past year of 13 week US Treasury Bills.  We apply floors to each index to avoid negative market returns.  We remove stocks with $\beta < 0$ or $\beta > 10$, and those without continuous trade data over the past year i.e. 253 days, in order to avoid market anomalies.  We select one-year data because we lose too much variability with five-year historical market data $\approx 2/3$.  We believe this is due to the market reverting to means.

Model adjustments during time of market turbulence:
Our model for stock selection requires mathematical adjustment in times of market declines when our indices move into negative territory over a year or when interest rates approach or drop below zero.  This impacts the signs of our scoring models.

\paragraph{Discussion of Quantum Advantage}
We optimize a reasonably sized portfolio using DWave’s 2,041 qubit quantum annealer through a repeatable research and business process. We pick 40 assets, which creates a solution space of $2^{40}$, or 1.1 trillion portfolios from which to select.  

As practitioners, our research indicates potential for quantum advantage at a higher number of assets. At lower asset levels there are efficient classical algorithms that do better.  We cannot solve this problem using brute force on our equipment at 40 assets.  We find an equivalent portfolio with a random sampling of 500M portfolios that takes hours to run.  One portfolio with 3 out of 40 assets appears optimal via a genetic algorithm seeded with 1,028 random portfolios.  We improve on that timing by seeding the genetic algorithm with the DWave solutions.

If we repeat the process of 36 experiments with 60 assets, we expect we might beat the genetic algorithm...but we would also use a classical simulated annealer which could outperform the genetic algorithm...so the race for quantum advantage continues.

\paragraph{Next Steps}
In the future we intend to evaluate reverse annealing, simulated annealing (both thermal and quantum), use of the DWave Hybrid solver, and to optimize larger and more diverse portfolios.  We also intend to further optimize the DWave runs and the QUBO build process.

From an economics perspective, we intend to add different types of financial assets, including bonds, commodities, real estate investment trusts, and currencies (including Bitcoin - USD).  We intend to evaluate additional economic factors such as financial health, growth, dividend payouts, and liquidity, which today we include implicitly.

Finally, we saw unique behavior when actual market returns over one year became negative (which it did during our research), and when individual stocks behaved erratically (e.g.,  $\beta < 0 $ and $\beta > 5.0$).  We intend to explore how these impact the portfolio optimization solutions from the Chicago Quantum Net Score.

\section{Thank You}

We acknowledge and thank the writers, maintainers and 
community contributors for Python, Numpy, Pandas, Matplotlib, Scipy, and DWave Ocean, Julia, and R.  We also thank DWave Systems, Google, Slack, Anaconda, and Jupyter for use of their tools in this research effort.  


\begin{thebibliography}{99}
	\bibitem{10Classes}{N.Bekkers, R.Doeswijk, T.Lam, Strategic Asset Allocation: Determining the Optimal Portfolio with Ten Asset Classes 
	\url{https://papers.ssrn.com/sol3/papers.cfm?abstract_id=1368689} Oct 20
	09}
	
    \bibitem{AltTabsen}{StackExchange, Comparing Portfolio Volatility with Index Volatility seems a wrong method?, Jan 2015, AltTabsen, QuantK, and Ric, 		\url{https://quant.stackexchange.com/questions/16223/comparing-portfolio-volatility-with-index-volatility-seems-a-wrong-method}}
	
	\bibitem{BETA}{BETA, Investopedia \url{https://www.investopedia.com/terms/b/beta.asp}}
	
	\bibitem{Brinson86}{G. Brinson, L.R. Hood, G. Beebower, Determinants of Portfolio Performance, June 1986, Financial Analysts Journal 42(4):39-44, 
	DOI: \url{10.2469/faj.v42.n4.39}}
	
	\bibitem{Brinson91}{G. Brinson, B. Singer, G. Beebower, Determinants of Portfolio Performance II: An Update, May/June 1991, 
		Financial Analysts Journal pp.40}
	
	\bibitem{Cholesky}{Ahmed Fasih, StackOverflow: Python: convert matrix to positive semi-definite, \url{https://stackoverflow.com/questions/43238173/python-convert-matrix-to-positive-semi-definite/43244194}}
	
	
	\bibitem{Divakar}{V. Divakar, Calculating the Covariance Matrix and Portfolio Variance, December 27, 2018, 
	\url{https://blog.quantinsti.com/calculating-covariance-matrix-portfolio-variance}}
	
	\bibitem{DwaveDocs}{DWave Systems Documentation,\url{https://www.docs.dwavesys.com}}
	
	\bibitem{DWaveRes}{DWave Systems Application Development Resources, \url{https://cloud.dwavesys.com/leap/resources}}
	
	
	\bibitem{EfficientFrontier}{Efficient Frontier at Wikipedia \url{https://en.wikipedia.org/wiki/Efficient_frontierEfficient Frontier Wiki}}
	
	\bibitem{Markowitz52}{Harry Markowitz, Portfolio Selection, The Journal of Finance, 7,1 (1952), 77-91}
	
	
	
	\bibitem{Marzec}{Michael Marzec. \emph{Portfolio Optimization: Applications in Quantum Computing.} SSRN Electronic Journal, 2013 \url{https://papers.ssrn.com/sol3/papers.cfm?abstract_id=2278729}}
	
	
	\bibitem{MachineLearning}{Marcos Lopez de Prado, Ph.D., Machine Learning Asset Allocation, \url{https://ssrn.com/abstract=3469964},
	Advanced in Financial Machine Learning, ORIE 5256}
	
	\bibitem{EF}{Nada Elsokkary, Faisal Shah Khan, Davide La Torre, Travis S. Humble, Joel Gottlieb. Financial Portfolio Management using D-Wave’s Quantum Optimizer: The Case of Abu Dhabi Securities Exchange. IEEE High Performance Extreme Computing Conference, 2017 (edited)}
	
	\bibitem{FRMwR}{Bernard Pfaff, Financial Risk Modelling and Portfolio Optimization with R, second edition, Wiley, 2016.  ISBN : 9781119119661}
	
	\bibitem{M_Jensen}{Michael Jensen, The Pricing of Capital Assets and the Evaluation of Investment Portfolios, The Journal of Business, Vol. 42, No.2,
		(Apr., 1969), pp. 167 - 247}
	
	\bibitem{Nielsen}{An Introduction to Information Geometry, \url{https://arxiv.org/pdf/1808.08271.pdf}}
	
	\bibitem{OML}{Orús, Román, Samuel Mugel, and Enrique Lizaso. Quantum computing for finance: overview and prospects. Reviews in Physics 4 (2019) (arXiv:1807.03890 [quant-ph])}
	
	\bibitem{ReSolve}{ReSolve Asset Management Whitepaper, A General Framework for Portfolio Choice, 27 pages, undated}
	
	\bibitem{Rubenstein}{M. Rubenstein, Markowitz's Portfolio Selection: a Fifty-Year Retrospective, The Journal of Finance, Vol LVII, Num 3, June 2002, DOI:10.1.1.404.4279}
	
	
	\bibitem{Sharpe}{Sharpe Ratio at Wikipedia \url{https://en.wikipedia.org/wiki/Sharpe_ratio}}	
	
	\bibitem{Sharpe94}{William F. Sharpe, Stanford University, The Sharpe Ratio, The Journal of Portfolio Management, Fall 1994
		\url{http://web.stanford.edu/~wfsharpe/art/sr/sr.htm}}
	
	\bibitem{SP}{Standard \& Poor, formerly CapitalIQ company database, Highland Park Public Library access to data screening tools,
		\url{https://hplibrary.org/databases-5317}}
	
	\bibitem{Sharpe_66}{William F. Sharpe, \emph{Mutual Fund Performance}, The Journal of B business, Vol 39, No.1, Part 2, Supplement on Security Prices.
		(Jan 1966), pp. 119 - 138 \url{http://www.jstor.org/stable/2351741}}
	
	
	\bibitem{Stockton}{Scot Stockton, \emph{What's The Difference Between 45 and 28 Percent Return? The Efficient Frontier}, August 20, 2018, Seeking Alpha,
		\url{https://seekingalpha.com/article/4200744-difference-45-return-and-28-efficient-frontier}}
			
	
	\bibitem{VK}{D.Venturelli, A.Kondratyev, \emph{Reverse Quantum Annealing Approach to Portfolio Optimization Problems}, arXiv:1810.08584, 2018}
	
	\bibitem{3Models}{X.Ye, L.Ning, J.Yang, D.Yao, Y.Hong, \emph{Research on three different Portfolio Models with singular Covariance Matrix} IOSR Journal of Mathematics (IOSR-JM) Volume 14, Issue 5 Ver. II (Sep - Oct 2018), PP 33-39}
	
		
	\bibitem{yfinance}{YFinance python module, maintained by Ran Aroussi, Thank you} \url{https://pypi.org/project/yfinance/}
	
	\bibitem{yahoof}{Yahoo Finance provides all historical data used in these experiments, Thank you, \url{https://finance.yahoo.com/}}
	

		
\end{thebibliography}
\end{document}